%% file: itw2010.tex
\documentclass[conference]{IEEEtran}

\usepackage[dvips]{graphicx}
\usepackage[tight]{subfigure}
\usepackage{amsmath}
\usepackage{amssymb}

\def \bm {\mathbf}

\newcounter{MYtempeqncnt}

\begin{document}

\title{Lossy Source Compression of Non-Uniform Binary Sources Using GQ-LDGM Codes}

\author{
\IEEEauthorblockN{Lorenzo Cappellari}
\IEEEauthorblockA{Dept. of Information Engineering, University of Padova, Italy\\
Email: lorenzo.cappellari@dei.unipd.it}
}

\maketitle

\begin{abstract}
\input{abstract.txt}
\end{abstract}

\section{Introduction}
In the proof of achievability of the channel capacity, as well as in the proof of achievability of the rate-distortion function, random ensembles of codes are tested and shown to be asymptotically optimal \cite{cover_ElemInfoThNew}. But, in practice, random codes are useless because without any structure no channel decoding/source encoding algorithms exist that run in reasonable time. In many cases of interest (e.g.~binary symmetric channels, binary source coding with Hamming-distance distortion measure, \dots) optimality is also achieved by linear codes. However, conducting a search among their codewords is still a tough task unless some other condition holds.

Luckily, \emph{sparse} structures, e.g.~linear codes with parity-check or generator matrices with a number of non-zero elements that is roughly proportional to the length of their codewords, have been shown to asymptotically perform very close to the theoretical bounds. Notably, for these structures, there often exist practical algorithms that essentially permit to reach their asymptotic performances.

For example, in the case of channel coding for the binary symmetric channel, the \emph{low density parity-check} (LDPC) codes are shown to almost achieve the channel capacity \cite{mackay_ldpc}, at least under maximum-likelihood decoding. In addition, they can be decoded in linear time with respect to the codeword length by using the standard message-passing algorithm \cite{kschischang_FactorGraphsAnd} with negligible performance losses \cite{richardson01_TheCapacityOf}.

On the other side, it has been proved that sparse codes are optimal too for lossy source coding. For example, \emph{low density generator matrix} codes (LDGM) almost achieve the rate-distortion function of the binary-symmetric memoryless source with Hamming-distance distortion measure \cite{wainwright10_LossySourceCompression%
}; also, by appropriately combining LDPC and LDGM codes it is possible to solve problems of coding with side information \cite{wainwright09_LowDensityGraph%
}. But, if the standard message-passing algorithm \cite{kschischang_FactorGraphsAnd} is employed in this dual setting, where there actually exist \emph{multiple} near-optimal encodings, it usually fails to converge or converges to meaningless marginals.

One of the first works showing that iterative algorithms can actually work as well for quantization appeared in \cite{martinian03_IterativeQuantization}, where the problem of \emph{binary erasure quantization} over LDGM codes was tackled. In order to solve the convergence problem, a simple modification of the message-passing algorithm was proposed in \cite{regalia09_AModifiedBelief} that led to satisfying results. Other works %
investigate adaptations of algorithms from the field of statistical physics in order to do data compression as well, e.g.~by using the \emph{survey propagation} algorithm \cite{mezard02_AnalyticAndAlgorithmic}. Furthermore, practical iterative algorithms that rely on \emph{non-linear} codes have been proposed for achieving the rate-distortion bound \cite{gupta09_NonlinearSparseGraph}.

One interesting possible workaround to achieve convergence in binary source coding with LDGM codes, as suggested in \cite{wainwright10_LossySourceCompression%
}, is to (i) run the traditional message-passing algorithm over an \emph{augmented} code-space where variables can also take a \emph{free} state, (ii) set the more certain variables to their preferred (non-free) states, and (iii) iterate this procedure over the \emph{decimated} code obtained by eliminating the variables that have been set in step (ii), until all variables have been set. Indeed, this approach can be seen as extending both the standard message-passing rules and the survey propagation ones \cite{maneva07_ANewLook}; in addition, it can be simplified as shown in \cite{filler07_BinaryQuantizationBP}.

In this paper we aim at extending the algorithms described in \cite{wainwright10_LossySourceCompression} to the case in which the binary source to be compressed is not uniform. In this case, in fact, a linear binary code whose codewords have a uniform statistics is no longer optimal \cite{cover_ElemInfoThNew}, as happens in the dual case of channel coding for non-symmetric channels. In the latter case optimality can still be achieved relying on non-binary LDPC codes %
and their variants, such as for example the \emph{$GF(q)$-quantized coset LDPC} (GQC-LDPC) codes \cite{bennatan04_OnTheApplication}. Among others, GQC-LDPC codes have interesting applications in the framework of channel coding with side information at the encoder%
, where they are applied in conjunction with good source codes in a superposed fashion \cite{bennatan_SuperpositionCodingFor}. In particular, in this paper, we define \emph{$GF(q)$-quantized LDGM} (GQ-LDGM) codes as duals of GQC-LDPC codes, and design practical algorithms for quantization over their codewords. GQ-LDGM codes can be eventually employed in conjunction with good channel codes for source coding with side information at the decoder%
, using a superposition scheme dual to the one in \cite{bennatan_SuperpositionCodingFor}.

As the question ``can linear codes with a suitable mapping function achieve the rate-distortion bound?'' was positively answered in \cite{chen07_AchievingTheRate}, and similar results have been proved for LDGM codes \cite{sun10_AchievingTheRate}, our focus on the practical side of the problem, namely on designing practical encoding algorithms, is fully justified. Other than this, constructions based on \emph{quantized} LDGM codes (or on \emph{multilevel quantization}) were used as well to tackle scalable coding \cite{zhang09_LdgmBasedCodes} and multiple description coding \cite{zhang09_MultipleDescriptionCoding} problems. This paper is also connected with \cite{gupta09_NonlinearSparseGraph}, where codes for non-uniform sources too are designed and tested.

The rest of this paper is organized as follows. In Section \ref{s:LDGM} we review the definition of $GF(q)$-LDGM codes and extend the space-augmentation procedure used in %
\cite{wainwright10_LossySourceCompression} to these codes. In Section \ref{s:GQ-LDGM} we introduce GQ-LDGM codes and describe the message-passing and decimation rules used for performing quantization. In Section \ref{s:results} we present some experimental results obtained with regular degree GQ-LDGM codes. Section \ref{s:concl} summarizes our conclusions.

\section{$GF(q)$-LDGM Codes and Their Augmentation}\label{s:LDGM}
An $(n,m)$ LDGM code over $GF(q)=\{l_0=0,l_1,\dots,l_{q-1}\}$ is easily specified by means of a \emph{factor graph} \cite{kschischang_FactorGraphsAnd} made by $n$ checks $a$ and $n$ \emph{constrained} variables $x_a$, $a \in \{0,1,\dots,n-1\}$, plus $m$ \emph{free} variables $z_i$, $i \in \{0,1,\dots,m-1\}$. Each check $a$ is connected to $x_a$ on one side and to $\bm{z} _{N(a)}$ on the other\footnote{$\bm{z}_{N(a)}$ stands for $\{z_i:i\in N(a)\}$.}, where $N(a) \subseteq \{0,1,\dots,m-1\}$; each such connection is labelled with a \emph{weight} $g_{ia}\in GF(q)\setminus \{l_0\}$. Each variable $z_i$ is then connected to the checks in $N(i) \subseteq \{0,1,\dots,n-1\}$. Variables take values on $GF(q)$, checks are satisfied iff $x_a = \sum_{i\in N(a)} g_{ia}z_i$, and the $(n+m)$-tuples $(\bm x,\bm z)$ such that all checks are satisfied form the codewords of the LDGM code.

\begin{figure}
\centering
  \includegraphics[scale=0.7]{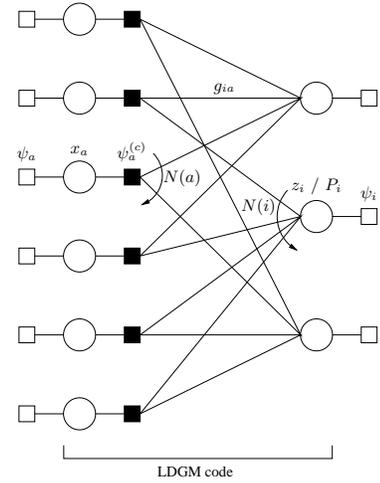}
  \caption{Factor graph of an LDGM code completed with the compatibility functions for assigning suitable probabilities to its generalized codewords.}
  \label{f:graph}
\end{figure}

The \emph{augmented LDGM} code is made of all the $(n+m)$-tuples of elements of $GF(q)^\ast \triangleq GF(q) \cup \{\ast\}$ such that, with reference to the factor-graph of the original code (see Fig.~\ref{f:graph}),
\begin{enumerate}
\item either checks connect to $x_a = \ast$ on one side and \emph{at least} one element in $\bm{z}_{N(a)}$ equals $\ast$ on the other, or they are satisfied (i.e.~their neighborhood is made of all elements of $GF(q)$ that satisfy the condition given above);
\item\label{l:cond} free variables different from $\ast$ connect to \emph{at least} two exactly satisfied checks (we assume that $|N(i)|\geq 2$).
\end{enumerate}
Those codewords are called \emph{generalized} codewords.

Due to the constraint in \ref{l:cond}, in order to check if an $(n+m)$-tuple is a generalized codeword, not only additional \emph{compatibility} functions must be imposed on the free variables, but also the corresponding nodes should be equipped with some knowledge about the state of the checks to which they connect. Similarly to what done in \cite{wainwright10_LossySourceCompression}, this can be easily achieved by replacing the nodes corresponding to the free variables with some variables $P_i \in \mathcal{P}_i \triangleq GF(q) \times \{1,2,\dots,2^{|N(i)|} - 1\} \cup \{\ast\}$, where $\{1,2,\dots,2^{|N(i)|} - 1\}$ is assumed in one-to-one correspondence with the non-empty sets in $\mathcal{P}\left(N(i)\right)$, the power set of $N(i)$. Once we define the two \emph{projections} $\pi : \mathcal{P}_i \rightarrow GF(q)^\ast$ and $\varphi : \mathcal{P}_i \rightarrow \mathcal{P}\left(N(i)\right)$ such that
\begin{eqnarray}
\pi(P_i) &=& \left\{
\begin{array}{ll}
\ast & ,\;\mbox{if } P_i = \ast \\
l_k & ,\;\mbox{if } P_i = (l_k, \cdot)
\end{array}\right. \nonumber\\
\varphi(P_i) &=& \left\{
\begin{array}{l@{\hspace{0pt}}l}
\emptyset & ,\;\mbox{if } P_i = \ast \\
A\subseteq N(i) & ,\;\mbox{if } P_i = (\cdot, \mbox{index of subset }A)
\end{array}\right.\;, \nonumber
\end{eqnarray}
and the compatibility functions at checks and nodes\footnote{$\chi[\cdot]$ is one iff the condition indicated is satisfied and zero otherwise; it is also assumed that $l_k+\ast=\ast+l_k=\ast$ and $\ast+\ast = \ast$.} \addtocounter{equation}{1} (\ref{e:check}) and 
\begin{equation}
\psi'_i(P_i) = \chi\left[ P_i = \ast \vee \left|\varphi(P_i)\right|\geq 2 \right]\;,
\end{equation}
the generalized codewords are all the $(n+m)$-tuples $\left(\bm x,\bm{\pi}(\bm P)\right)$ such that all $x_a$ and $P_i$ satisfy the new compatibility functions. In practice, $P_i$ brings both the value of the free variable, and, in case of the value being different from $\ast$, the indices of the checks that are enforcing this value (note that at least one such check always exists).

\begin{figure*}[!bt]
\normalsize
\setcounter{MYtempeqncnt}{\value{equation}}
\setcounter{equation}{2}
\begin{eqnarray}
\psi_a^{(c)}(\bar{x}_a,\bm{P}_{N(a)}) &=& \chi\left[
\left(\sum_{i\in N(a)}\pi(P_i) = \ast \wedge \bar{x}_a = \ast \wedge \bigwedge_{i\in N(a)} x_a\notin\varphi(P_i)\right) \right. \vee \label{e:check}\\
&& \vee \left.
\left(\sum_{i\in N(a)}\pi(P_i) \ne \ast \wedge \bar{x}_a = \sum_{i\in N(a)}g_{ia}\pi(P_i) \wedge \bigwedge_{i\in N(a)} x_a\in\varphi(P_i) \right)\right] \nonumber
\end{eqnarray}
\setcounter{equation}{\value{MYtempeqncnt}}
\hrulefill
\vspace*{2pt}
\end{figure*}

\section{Binary Source Compression with $GF(q)$-quantized LDGM codes}\label{s:GQ-LDGM}
At this point, when one wants to use the original LDGM code in a quantized fashion for quantizing a random realization of a binary codeword, a map $Q : GF(q)^\ast \rightarrow GF(2)^\ast$ is introduced such that $Q(x_a)=\ast$ iff $x_a=\ast$. This map eventually converts a $GF(q)$ codeword corresponding (through standard LDGM encoding) to a compressed representation in $GF(q)$ into the $GF(2)$ codeword used as source reconstruction.

Then, by adding the compatibility functions for the constrained variables $\psi_a(\bar{x}_a;s_a,\beta,w_s) = \psi^{(Q)}_a\left(Q(\bar{x}_a);s_a,\beta,w_s\right)$, where
\begin{equation}
\psi^{(Q)}_a(x^{(Q)}_a;s_a,\beta,w_s) = \left\{\begin{array}{ll}
  e^\beta    & ,\;\mbox{if } x^{(Q)}_a = s_a \\
  w_s        & ,\;\mbox{if } x^{(Q)}_a = \ast \\
  e^{-\beta} & ,\;\mbox{if } x^{(Q)}_a = s_a + 1
\end{array}\right.\;,
\end{equation}
and by using $\psi_i(P_i;w_i) = w_i^{\chi[P_i = \ast]}\psi'_i(P_i)$ as compatibility functions for the free variables, a probability distribution over all $(n+m)$-tuples $(\bm{x},\bm{P})$ is enforced. This distribution, when marginalized over all the $(\bm{x},\bm{P})$ tuples corresponding to the same generalized codeword $(\bm{x},\bm{z})$, leads to a probability
\begin{equation}
P[(\bm{x},\bm{z});s_a,\beta,w_s,w_i] \propto w_i^{N^\ast(\bm{z})}\cdot w_s^{N^\ast(\bm{x})}\cdot e^{-2\beta d_H\left(Q(\bm{x}),\bm{s}\right)}\;.
\end{equation}

The factor graph describing this distribution is shown in Fig.~\ref{f:graph}. In the formulas, $\bm s$ denotes the $n$-dimensional source realization to be quantized, $\beta\geq 0$ determines how much more probability should be given to the values (taken by constrained variables) that after quantization match the source with respect to those that will not match it, and $w_s\geq 0$ and $w_i\geq 0$ determine how much each value $\ast$ (occurring in constrained or free variables, respectively) modifies the overall generalized codeword probability. In particular, $w_s> 1$ (or $w_i> 1$) will make the $\ast$ value to be preferred with respect to symbols in $GF(q)$. Also, $N^\ast(\cdot)$ denotes the number of $\ast$ values in a given tuple of $GF(q)^\ast$, and $d_H(\cdot,\cdot))$ denotes the Hamming distance between tuples of $GF(2)^\ast$, assuming that $\ast$ has distance $1/2$ from any other symbol.

Despite the exponentially large cardinality of $\mathcal{P}_i$, that in practice seems to prevent the utilization of message-passing algorithms, it can be shown that the messages over each edge can be described by only $q+2$ probabilities, at least in the case where $q$ is a \emph{prime} number. In particular, in the check-to-variable direction these probabilities are $\mu_{ai}^\ast$, $\mu_{ai}^{\neq}$, and $\mu_{ai}^{(l_k)}$, for $k=0,1,\dots,q-1$, because the messages satisfy
\begin{eqnarray}
\mu_{ai}\left(P_i=\ast\right) &\equiv& \mu_{ai}^\ast \\
\mu_{ai}\left(P_i:P_i\neq \ast \wedge x_a\notin\varphi(P_i) \right) &\equiv& \mu_{ai}^{\neq} \\
\mu_{ai}\left(P_i:\pi(P_i) = l_k \wedge x_a\in\varphi(P_i) \right) &\equiv& \mu_{ai}^{(l_k)}\;.
\end{eqnarray}
In the variable-to-check direction, it is similarly sufficient to specify the probabilities $\mu_{ia}^\ast$, $\mu_{ia}^{\neq}$, and $\mu_{ia}^{(l_k)}$, for $k=0,1,\dots,q-1$, which are defined as
\begin{eqnarray}
\mu_{ia}(P_i = \ast) &\triangleq& \mu_{ia}^\ast \\
\sum_{P_i:P_i\neq \ast \wedge x_a\notin\varphi(P_i)} \mu_{ia}(P_i) &\triangleq& \mu_{ia}^{\neq} \\
\sum_{P_i:g_{ia}\pi(P_i)=l_k \wedge x_a\in\varphi(P_i)} \mu_{ia}(P_i) &\triangleq& \mu_{ia}^{(l_k)}\;.
\end{eqnarray}

With these definitions, the messages from check $a$ to variable $i$ can be computed by
\begin{eqnarray}
\mu_{ai}^\ast &=& w_s \prod_{j\in N(a)\setminus \{i\}} \left[\mu_{ja}^\ast + \mu_{ja}^{\neq} \right]\\
\mu_{ai}^{\neq} &=& w_s \left[ \prod_{j\in N(a)\setminus \{i\}} \left[\mu_{ja}^\ast + \mu_{ja}^{\neq} \right] - \prod_{j\in N(a)\setminus \{i\}} \mu_{ja}^{\neq} \right]\nonumber\\
\mu_{ai}^{(l_k)} &=& \sum_{r=0}^{q-1}\psi_a(l_r) \lambda_{ai}^{(l_r-g_{ia}l_k)}\;,
\end{eqnarray}
where\footnote{$[\mu_{ja}^{(l_k)}]$ denotes the column vector whose $q$ components are obtained for $k=0,1,\dots,q-1$; also, components of product of vectors are assumed to be products of the corresponding components. DFT$_q$ and IDFT$_q$ denote $q$-point Fourier transformation and its inverse, respectively. As a remark, we were able to derive the given formula for $[\lambda_{ai}^{(l_k)}]$ only for prime values of $q$.}
\begin{equation}
[\lambda_{ai}^{(l_k)}] = \mbox{IDFT}_q\left[ \prod_{j\in N(a)\setminus \{i\}} \mbox{DFT}_q[\mu_{ja}^{(l_k)}] \right]\;.
\end{equation}
The messages from variable $i$ to check $a$ can be instead computed by
\begin{eqnarray}
\mu_{ia}^\ast &=& w_i \prod_{c\in N(i)\setminus \{a\}} \mu_{ci}^\ast\\
\mu_{ia}^{\neq} &=& \sum_{k=0}^{q-1} \prod_{c\in N(i)\setminus \{a\}} \left[ \mu_{ci}^{\neq} + \mu_{ci}^{(l_k)} \right] - q \prod_{c\in N(i)\setminus \{a\}} \mu_{ci}^{\neq} - \nonumber\\
&-& \sum_{b\in N(i)\setminus \{a\}} \left[ \sum_{k=0}^{q-1} \mu_{bi}^{(l_k)} \right] \prod_{c\in N(i)\setminus \{a,b\}} \mu_{ci}^{\neq}\\
\mu_{ia}^{(l_k)} &=& \prod_{c\in N(i)\setminus \{a\}} \left[ \mu_{ci}^{\neq} + \mu_{ci}^{(g_{ia}^{-1}l_k)} \right] - \prod_{c\in N(i)\setminus \{a\}} \mu_{ci}^{\neq}
\end{eqnarray}

Finally, the marginals with respect to each $z_i$ are found using the following equations:
\begin{eqnarray}
P[z_i = \ast] \triangleq p_i^\ast &=& w_i \prod_{a\in N(i)} \mu_{ai}^\ast\\
P[z_i = l_k] \triangleq p_i^{(l_k)} &=& \prod_{a\in N(i)} \left[\mu_{ai}^{\neq} + \mu_{ai}^{(l_k)} \right] - \prod_{a\in N(i)} \mu_{ai}^{\neq} -\nonumber\\
&-& \sum_{b\in N(i)} \mu_{bi}^{(l_k)} \prod_{a\in N(i)\setminus \{b\}} \mu_{ai}^{\neq}\;,
\end{eqnarray}
where $k=0,1,\dots,q-1$.

As a remark, the derivation of these message-passing rules is quite straightforward, despite of combinatorial nature. Due to the limited space, the actual derivations were not included; however, one can notice how these message-updating rules appear as an extension of the ones relative to the $GF(2)$ field and derived in \cite{wainwright10_LossySourceCompression}.

In practice, the quantization algorithm we designed follows the following steps:
\begin{enumerate}
  \item\label{en:init} All check-to-variable messages are initialized by assuming that there are no incoming variable-to-check messages; in particular, this implies $\mu_{ai}^\ast = w_s$, $\mu_{ai}^{\neq} = 0$, and $\mu_{ai}^{(l_k)} = \psi_a(g_{ia}l_k)$;
  \item free variables that are either disconnected from any check or that connect exclusively to degree one checks are \emph{removed} from the factor-graph;
  \item message-passing rules for computing variable-to-check messages and recomputing check-to-variable messages are iteratively applied until convergence is achieved (or a maximum number of iterations is reached);
  \item free variables that are sufficiently \emph{biased} are \emph{removed} from the factor-graph;
  \item if there are still variables to be \emph{removed}, go back to step \ref{en:init}; otherwise exit the algorithm.
\end{enumerate}
Each time a free variable $z_i$ is \emph{removed}, its value is set as the $l_k$ in $GF(q)$ that maximizes $p_i^{(l_k)}$; edges connecting that variable to checks are removed as well, and the compatibility functions $\psi_a$, for each $a\in N(i)$, are updated in order to take into account for this removal. Of course, unconnected checks are removed too from the factor-graph.

In addition, in order to quantize the \emph{bias} of a probability distribution over $GF(q)^\ast$, we used the \emph{unbiased} standard deviation of the vector $[p_i^{(l_k)}]$ multiplied by $\sqrt{q}$. In particular, if $\sum_{k=0}^{q-1}p_i^{(l_k)}=S$, i.e.~$p_i^\ast=1-S$, it turns out that the bias $B_i$ equals
\begin{equation}
B_i = S\sqrt{\frac{q\sum_{k=0}^{q-1}\left(\frac{p_i^{(l_k)}}{S}\right)^2 - 1}{q - 1}}\;,
\end{equation}
which is always in the range $0\leq B_i\leq 1$. As a remark, note that the maximum is achieved iff $S=1$ (i.e.~$p_i^\ast=0$) and exactly one element of $[p_i^{(l_k)}]$ equals one. This formulation extends the bias concept introduced in \cite{wainwright10_LossySourceCompression}.

\section{Experimental Results}\label{s:results}
In the experiments we defined the quantizer $Q(\cdot)$ on the base of a parameter $0<Q_m<q$, that specifies the minimum value in $GF(q)$ to be quantized as the non-zero value of $GF(2)$. In practice, $Q(l_k)=l_0$ for all $k<Q_m$, while $Q(l_k)=l_1$ otherwise. Hence, the employed GQ-LDGM codes have elements assuming the value $l_1$ with probability $r=\frac{q-Q_m}{q}$. We tested $4$ different combinations of $q$ and $Q_m$, as shown in Table \ref{t:exp}.

\begin{table}
\caption{Degrees, rates and source probability used in the various experiment sets.}
\label{t:exp}
\centering
\begin{tabular}{c|c|c|c|c}
\hline
Experiment set & $1$ & $2$ & $3$ & $4$ \\
\hline
$q$   & $5$     & $3$     & $5$     & $2$ \\
$Q_m$ & $4$     & $2$     & $3$     & $1$ \\
$r$   & $0.200$ & $0.333$ & $0.400$ & $0.500$ \\
$d_c$ & $2$     & $2$     & $2$     & $2$ \\
$d_v$ & $9$     & $6$     & $9$     & $4$ \\
$R$ [bit/sample] & $0.516$ & $0.528$ & $0.516$ & $0.500$ \\
$p_s$ & $0.230$ & $0.365$ & $0.420$ & $0.500$ \\
\hline
\end{tabular}
\end{table}

In these preliminary experiments, we derived the GQ-LDGM codes from randomly generated $GF(q)$-LDGM codes with regular degree distributions. In particular, checks and free variables have degree $d_c$ and $d_v$ respectively, so that the resulting (source coding) rate, expressed in bit/sample units, equals $R = \frac{d_c}{d_v}\log_2(q)$. We fixed the check degree $d_c = 2$ and derived, for each employed value of $q$, a suitable value of $d_v$ for obtaining a rate of approximately $0.5$ bit/sample, as shown in Table \ref{t:exp}.\footnote{As noted in \cite{wainwright10_LossySourceCompression} and \cite{regalia09_AModifiedBelief}, in the case of irregular degree distributions it is important to have an high ratio of checks of degree $2$. This is the reason why we chose $d_c = 2$ in our experiments with regular distributions. Note that this choice, in case of $q=2$, introduces linear dependency between the rows of the LDGM generating matrix. In practice, the actual rate could be (very slightly) reduced as there always exist $2$ $m$-tuples of free-variables leading to the same LDGM codeword.}

In order to use GQ-LDGM codeword statistics that match the optimal distribution, i.e.~the one achieving the theoretical rate-distortion bound at the employed rates, in each experiment we generated a random i.i.d.~source in $GF(2)$ with a probability of symbol $l_1$ equal to the values of $p_s$ shown in Table \ref{t:exp}. These values are such that, approximately,
\begin{equation}
\frac{p_s - D(R; p_s)}{1 - 2D(R; p_s)} = r\;,
\end{equation}
where $D(R; p_s)$ is the theoretical (Hamming) distortion at the employed rates $R$, according to the Bernoulli distortion-rate function \cite{cover_ElemInfoThNew}. In particular, $D$ satisfies $H(D) = H(p_s) - R$, where $H(\cdot)$ is the entropy of the binary distribution with a given symbol probability.

In all experiments we assumed that the message-passing procedure relative to the various decimation steps converged when the maximum absolute difference between the (normalized) marginals $[p_i^\ast;p_i^{(l_k)}]$ computed in the current iteration and the ones computed in the previous one is less than $MP_{th}=0.05$, for all free-variables. Also, we fixed the maximum number of iterations at $MP_{\max}=100$, which was typically reached in the first couple of decimation steps.

Similarly to what done in \cite{wainwright10_LossySourceCompression}, after each message-passing run we removed at least $r_m=1$\% of the variables (starting from the more biased ones), as well as all the variables with a bias greater or equal than $B_m=0.7$, until a maximum of $r_m=10$\% of the variables was removed. We also borrowed from \cite{wainwright10_LossySourceCompression} the choices of $w_i=\exp(0.05)$ and $w_s=\exp(0.10)$. Instead, in order to suitably choose $\beta$ accordingly to the employed rate (and source statistics), we propose to choose it in a way such that the compatibility function $\psi_a(x_a)$ corresponds to the likelihood function of the \emph{test channel} \cite{cover_ElemInfoThNew} related to the source coding problem at hand\footnote{This is reasonable as we may assume that the source (with probability $p_s$) is in fact a corrupted version of a codeword (with probability $r$), obtained after the latter went through the test channel (with probability of error $D(R; p_s)$).}; in practice, we chose
\begin{equation}
\beta=\frac{1}{2}\log\frac{1 - D(R; p_s)}{D(R; p_s)}\;,
\end{equation}
which equals approximately $1.53$, $1.19$, $1.10$, and $1.05$, respectively for each experiment set.

\begin{figure}
\centering
  \includegraphics[width=\linewidth]{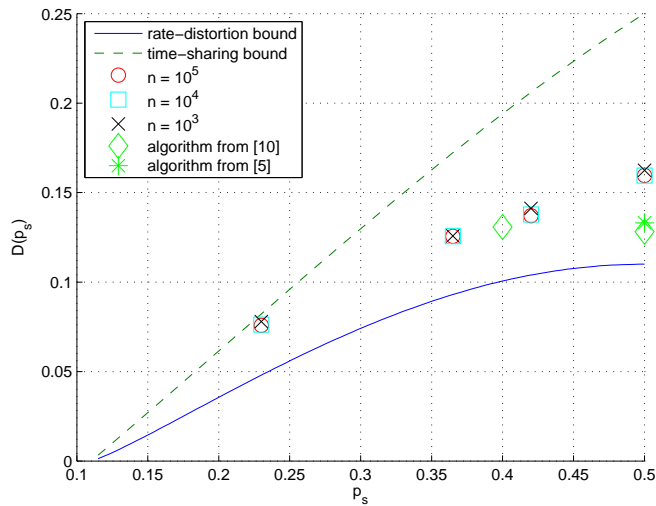}
  \caption{Empirical performance at $R\cong 0.5$ bit/sample compared with the rate-distortion and time-sharing bounds and with some data from \cite{gupta09_NonlinearSparseGraph}.}
  \label{f:res}
\end{figure}

The results, reported in Fig.~\ref{f:res}, are relative to compression of random realizations of length $n=10^5$, $10^4$ and $10^3$; they have been averaged over $4$ GQ-LDGM codes with $100$ compressions/code, $10$ codes with $500$ compressions/code, and $10$ codes with $1000$ compressions/code, respectively for the three lengths. From the plot, where the rate-distortion bound $H(D) = H(p_s) - R$ and the time-sharing bound (between the zero-rate and the zero-distortion points)
\begin{equation}
  D=p_s\frac{H(p_s)-R}{H(p_s)}
\end{equation}
are given too, we concluded that the proposed algorithm performed quite close to the rate-distortion bound, and always better than the time-sharing bound. Surprisingly, the performance turned out to be essentially independent from the codeword length.

The algorithm was implemented in C and was run on an Intel Pentium 4 CPU operating at $3$ GHz, with $1.5$ GByte of RAM. In terms of execution speed it turned out to be very competitive with respect to the timings given in \cite{wainwright10_LossySourceCompression}. In fact, in the experiments of set $1$, $80.53$, $6.96$ and $0.52$ seconds were necessary on average for performing each source compression, respectively for realizations of length $n=10^5$, $10^4$ and $10^3$. In the experiments of set $2$-$4$ the timings are $53.27$, $3.98$, and $0.31$; $78.70$, $6.93$, and $0.54$; and $46.74$, $2.28$, and $0.18$, respectively.

\section{Conclusion}\label{s:concl}
In this paper we presented a novel message-passing/decimation algorithm for binary source quantization. The algorithm is based on the utilization of $GF(q)$-quantized LDGM codes, that permit to generate codewords with a non-uniform distribution. Eventually, this allows to directly compress sources with non-uniform distribution too. Among the applications, the proposed codes could be used in conjunction with good channel codes in problems of source coding with side information at the decoder.

The compression performances, obtained with regular degree distributions, show that the proposed algorithms perform reasonably close to the rate-distortion bound and always better than the time-sharing bound. It is also reasonable to assume that better performances could be achieved by suitably designing irregular degree-distribution codes, by augmenting the field cardinality $q$, or by employing damping strategies similar to the ones in \cite{filler07_BinaryQuantizationBP} for solving convergence problems, which are possible directions for future research on this subject.

\bibliographystyle{IEEEtran}
\bibliography{IEEEabrv,nonIEEEabrv,refs_books,refs_DSC,refs_other}

\end{document}

%% file: abstract.txt
In this paper, we study the use of GF(q)-quantized LDGM codes for binary source coding. By employing quantization, it is possible to obtain binary codewords with a non-uniform distribution. The obtained statistics is hence suitable for optimal, direct quantization of non-uniform Bernoulli sources. We employ a message-passing algorithm combined with a decimation procedure in order to perform compression. The experimental results based on GF(q)-LDGM codes with regular degree distributions yield performances quite close to the theoretical rate-distortion bounds.